\documentstyle[12pt,axodraw]{article}

\input epsf
\newcommand{\grpicture}[1]
{
    \begin{center}
        \epsfxsize=200pt
        \epsfysize=0pt 
        \vspace{-5mm}
        \parbox{\epsfxsize}{\epsffile{#1.ps}} 
        \vspace{5mm}
    \end{center}
}
\newcommand{\be}{\begin{eqnarray}}
\newcommand{\ee}{\end{eqnarray}}

\renewcommand{\vec}[1]{{\bf #1}}
\begin{document}
\begin{flushright}
Preprint ITEP -- 10/96 -TH
\end{flushright}

\vspace{2cm}

\begin{center}
{\Large Physics of Quark--Gluon Plasma}
\footnote{Lecture at the XXIV ITEP Winter School
(Snegiri, February 1996).}
\vspace{1cm}

A.V. Smilga\\
\vspace{0.5cm}

{\it ITEP, B. Cheremushkinskaya 25, Moscow 117259, Russia}
\vspace{1cm}
\end{center}

\abstract{In this lecture, we give a brief review of
what theorists now know, understand, or guess about static
and kinetic properties of quark--gluon plasma. A particular
attention is payed to the problem of physical observability,
i.e. the physical meaningfulness of various characteristics
of $QGP$ discussed in the literature.}

\section{Introduction.}
 It is well-known {\it for theorists} that the physics of
$QCD$ medium at high temperatures $T \gg \Lambda_{QCD}$
differs dramatically from how the system behaves at zero and
at low temperatures. At low temperatures, we have
confinement and the spectrum involves colorless hadron
states. The lightest states are pions --- pseudogoldstone
states which appear due to spontaneous breaking of
(approximate) chiral symmetry of $QCD$ lagrangian.

At high temperatures, hadrons get ``ionized'' to quarks and
gluons, and, in the $0^{\underline{th}}$ approximation, the
system presents the heat bath of freely propagating colored
particles. For sure, quarks and gluons interact with each
other, but at high temperatures the effective coupling
constant is small $\alpha_s(T) \ll 1$ and the effects due to
interaction can be taken into account perturbatively
\footnote{We state right now, not to astonish the experts,
that there are limits of applicability of perturbation
theory even at very high temperatures, and we {\it are}
going to discuss them later on.}.
This interaction has the long--distance Coulomb nature,
and the properties of the system are
in
many respects very similar to the properties of the usual
non-relativistic plasma involving charged particles with
weak
Coulomb interaction. The only difference is that quarks and
gluons carry not the electric, but color charge. Hence the
name: {\it Quark-Gluon Plasma} ($QGP$).

To describe the properties of $QGP$ is a rather interesting
problem of theoretical physics lying on cross-roads between
the relativistic field theory and condensed matter physics.
Personally, I had a great fun studying it. Unfortunately, at
present only a theoretical study of the problem is possible.
Hot hadron medium with the temperature above phase
transition can be produced for tiny fractions of a moment in
heavy ion collisions but:
 \begin{itemize}
\item It is not clear at all whether a real thermal
equilibrium is achieved.
\item A hot system created in the collision of heavy
nuclei rapidly expands and cools down emitting pions. It is
not possible to probe the properties of the system directly,
but only indirectly via the characteristics of the final
hadron state.
\item Anyway,  the temperature
achieved at existing accelerators is not high enough for the
perturbation
theory to
work and there are no {\it quantitative} theoretical
predictions with which experimental data can be compared.
May be RHIC would be better in this respect.
  \end{itemize}

Thus, at present, there are no experimental tests of
nontrivial theoretical predictions for $QGP$ properties. The
effects observed in experiment such as the famous
$J/\psi$ -- suppression (for a recent review see \cite{Jpsi}) 
just indicate that a
hot and dense medium is created but says little on whether
it is $QGP$ or something else.

The absence of feedback between theory and experiment is a
sad and unfortunate reality of our time: generally, what is
interesting theoretically is not possible to measure and
what is possible to measure is not interesting theoretically
\footnote{One of the possible exceptions of the general rule
in the field of thermal QCD is  a fascinating perspective to
observe the phenomenon of disoriented chiral condensate at
RHIC. But that refers to the temperature region $T \sim T_c
\sim 200 Mev$ where the phase transition with restoration of
chiral symmetry occurs (for a review of physics of phase
transition in QCD, see e.g. \cite{Varenna}).}.

Thus, we will not attempt in this lecture to establish
relation of the results of theoretical calculations with
realistic accelerator experiments. What we {\it will} do,
however, is discussing the relation of theoretical results
with {\it gedanken} experiments. Suppose, we have a thermos
bottle with $QGP$ on a laboratory table and are studying it
from {\it any} possible experimental angle.
We call a quantity physical if it can in principle be
measured in such a study and non-physical otherwise. We
shall see later that many quantities discussed by
 theorists may be called physical only with serious
reservations, and some are  not physical at all.

\section{Remarks on finite $T$ diagram technique.}
\setcounter{equation}0
 The main point of interest of these notes are the physical
phenomena in hot QCD system. However, as the main
theoretical tool to study them is the perturbation theory
and we want in some cases not only to quote the results, but
also to explain how they are obtained, we are in a position
to
describe briefly how the perturbative calculations at finite
$T$ are performed.

There are two ways of doing this --- in imaginary or in real
time. These techniques are completely equivalent and  which
one to use is mainly a matter of taste. Generally, however,
the Euclidean technique is more handy when one is interested
in pure static properties of the system (thermodynamic
properties and static correlators) where no real time
dependence is involved. On the other hand, when one is
interested in kinetic properties (spectrum of collective
excitations, transport phenomena, etc.), it is more
convenient
to calculate directly in real time.

\subsection{Euclidean (Matsubara) technique.}
 Consider a theory of real scalar field described by the
hamiltonian $H[\phi(\vec{x}),\  \Pi(\vec{x})]$. The
partition
function of this theory at temperature $T$ can be written as

\be
\label{Zdef}
Z = {\rm Tr} \left\{e^{-\beta H} \right\} =
\int \prod_\vec{x} d\phi(\vec{x}) {\cal K} [\phi(\vec{x}),
\phi(\vec{x}) ; \beta]
\ee
where ${\cal K}$ is the quantum evolution operator in the
imaginary time $\beta = 1/T$ :
 \be
\label{K}
 {\cal K} [\phi'(\vec{x}), \phi(\vec{x}) ; \beta]
= \sum_n   \Psi_n^*[\phi'(\vec{x})]  \Psi_n[\phi(\vec{x})]
e^{-\beta E_n}
\ee
$\Psi_n$ are the eigenstates of the hamiltonian. One can
express the integral in RHS of Eq.(\ref{Zdef}) as an
Euclidean path integral:

\be
\label{Zpath}
Z = \int \prod_{\vec{x}, \tau} d\phi(\vec{x}, \tau)
\exp \left\{ - \int_0^\beta d\tau \int d\vec{x} \ {\cal L}
[\phi(\vec{x}, \tau] \right\}
\ee
where the periodic boundary conditions are imposed
  \be
  \label{bc}
 \phi(\vec{x}, \tau + \beta) = \phi(\vec{x}, \tau)
 \ee
A thermal average $<{\cal O}>_T$ of any operator ${\cal O}
[\phi(\vec{x}, \tau)]$ has the form \cite{Mats}

\be
\label{OT}
<{\cal O}>_T = Z^{-1}
\int \prod_{\vec{x}, \tau} d\phi(\vec{x}, \tau) {\cal O}
[\phi(\vec{x}, \tau)]
\exp \left\{ - \int_0^\beta d\tau \int d\vec{x} \ {\cal L}
[\phi(\vec{x}, \tau] \right\}
\ee

One can develop now the diagram technique in a usual way.
The only difference with the zero temperature case is that
the Euclidean frequencies of the field $\phi(\vec{x}, \tau)$
are now quantized due to periodic boundary conditions
(\ref{bc}):
\be
\label{p0bos}
p_0^n = 2\pi i nT
\ee
with integer $n$. To calculate something, one should draw
the same graphs as at zero temperature and go over into
Euclidean space where the integrals over Euclidean
frequencies are substituted by sums:
 \be
 \label{intT}
\int \frac {d^4p}{(2\pi)^4} f(p) \longrightarrow
T \sum_n \int \frac {d^3p}{(2\pi)^3} f(2\pi i nT, \vec{p})
\ee
The same recipe holds in any theory involving bosonic
fields. In theories with fermions, one should impose
antiperiodic boundary conditions on the fermion fields
$\psi(\vec{x}, \tau)$ (see e.g. \cite{Brown} for  detailed
pedagogical explanations), and the frequencies are quantized
to
\be
\label{p0ferm}
p_0^n = i\pi (2n+1) T
\ee
An important heuristic remark is that, when the temperature
is very high, in many cases only the bosons with zero
Matsubara frequencies $p_0 = 0$ contribute in $<{\cal
O}>_T$. The contribution of higher Matsubara frequencies and
also the contribution of fermions in $<{\cal O}>_T$ become
irrelevant and, effectively,  we are dealing with a 3-
dimensional theory. As was just mentioned, it is true in
many, but not in {\it all} cases. For example, it makes no
sense to neglect fermion fields when one is interested in
the properties of collective excitations with fermion
quantum numbers. For any particular problem of interest a
special study is required.

\subsection{Real time (Keldysh) technique.}
Matsubara technique is well suited to find thermal averages
of {\it static} operators ${\cal O}(\vec{x})$. If we are
interested in a time-dependent quantity, there are two
options: {\bf i)} To find first the thermal average $<{\cal
O}(\vec{x}, \tau)>$ for Euclidean $\tau$ and perform then
an analytic continuation onto the real time axis. It is
possible, but quite often rather cumbersome.
{\bf ii)} To work in the real time right from the beginning.
The corresponding technique was first developed in little
known papers \cite{Bakshi} and independently by Keldysh
\cite{Keld} who applied it to condensed matter problems with
a particular emphasis on the systems out of thermal
equilibrium. It was fully apprehended by experts in
relativistic field theory only in the beginning of nineties.

As we do not really need the full--scale Keldysh technique in what
follows, we will not discuss it here, referring a reader to
the textbook \cite{LP}, to a hard-to-read, but
extensive review \cite{Landsman}, and to our papers
\cite{spectr,andam,damp1} where the real time technique was
applied for studying the properties of $QGP$. We only
mention
here that at finite T one should distinguish carefully
between the retarded,  advanced and mixed components of the
Green's functions which can be ``organized'' in a $2 \times
2$ matrix.

It suffices often to use a simplified version of the real
time technique due to Dolan and Jackiw \cite{DJ}. Let us
find the tree propagator of  real scalar field at finite
temperature:
  \be
  \label{propdef}
<\phi(x) \phi(0)>_T = \sum_n e^{- \beta E_n} <n|\phi(x)
\phi(0) |n>
  \ee
Introduce as usual a finite spatial volume $V$ and decompose
\be
\label{second}
\phi(x) = \sum_\vec{p} \frac 1{\sqrt{2\omega V}}
\left[ a_\vec{p} e^{-i \omega t + i \vec{p}\vec{x}}
+  a^+_\vec{p} e^{i \omega t - i \vec{p}\vec{x}} \right]
\ee
where $a^+_\vec{p}$ and  $a_\vec{p}$ are the creation and
annihilation operators. At zero temperature only the vacuum
average contributes and we can use the fact $a_\vec{p}|0> =
0$ to obtain a usual expression for the propagator. At
finite temperature, the excited states $|n>$ with
$a_\vec{p}|n> \neq 0$
contribute in the sum
(\ref{propdef}), and an
additional contribution to the propagator arises. We obtain
  \be
  \label{DJ}
\int <\phi(x) \phi(0)>_T e^{ipx} = \frac i{p^2 - m^2 + i0}
+ 2\pi \delta(p^2-m^2) n_B(|p_0|)
 \ee
where
  \be
  \label{nB}
n_B(\epsilon) = \frac 1{e^{\beta \epsilon} -1}
  \ee
is the Bose distribution function. An additional thermal
contribution
reflects the presence of real bosons on mass shell in the
heat bath. A fermion propagator is
derived in a similar way  and an additional  contribution is
proportional to
the Fermi distribution
function $n_F(\epsilon) = 1/(e^{\beta\epsilon} + 1)$.

The recipe is to substitute the Dolan-Jackiw propagators
(and in cases when it does not work --- the full-scale
matrix
Keldysh propagators) in the loop integrals for the usual
Feynman graphs. The integrals are done now over {\it real}
frequencies.

\section{Static Properties of $QGP$: a Bird Eye's View.}
\setcounter{equation}0

\subsection{Thermodynamics}

The basic thermodynamic characteristic of  a finite $T$
system is its free energy. We have in the lowest order for a
pure gluon system

\be
\label{Fgdef}
F^g = -T \ln Z = -T \ln \left[ \prod_\vec{p} \left(
\sum_{n=0}^\infty e^{-\beta n |\vec{p}|} \right)^{2(N_c^2-
1)} \right]
\ee
where $2(N_c^2-1)$ is the number of degrees of freedom of
the gluon field (the factor 2 comes due to two
polarizations) and the sum $\sum_{n} e^{-\beta n |\vec{p}|}$
is the free energy of a single boson field oscillator with
the frequency $\omega = |\vec{p}|$. Trading the sum for the
integral, we obtain for the volume density of the free
energy
 \be
\label{Fg0}
 \frac{F^g}{V} = 2(N_c^2 -1) T \int \frac {d^3p}{(2\pi)^3}
\ln \left[ 1 - e^{-\beta |\vec{p}|} \right] =
- \frac{\pi^2 T^4}{45} (N_c^2 -1)
\ee
 This is nothing else as the Stefan--Boltzman formula
multiplied by the color factor $N_c^2-1$. The quark
contribution is obtained quite similarly:
 \be
\label{Fqdef}
F^q = -T \ln Z = -T \ln \left[ \prod_\vec{p} \left( 1 + e^{-
\beta n |\vec{p}|} \right)^{4N_c N_f} \right]
\ee
where $4N_c N_f$ is the number of degrees of freedom and the
Pauli principle is taken into account. We obtain
 \be
\label{Fq0}
 \frac{F^q}{V} \ = \ - 4N_c N_fT \int \frac {d^3p}{(2\pi)^3}
\ln \left[ 1 + e^{-\beta |\vec{p}|} \right] \ =\ 
 - \frac{7\pi^2 T^4}{180} N_c N_f 
\ee
All other thermodynamic quantities of interest can be
derived
from the free energy by standard thermodynamic relations.
For example, the pressure just coincides with the free
energy with the sign reversed. The energy density is
  \be
 \label{Edef}
E = F - T \frac {\partial F}{\partial T}
 \ee
For massless particles in the lowest order the relation
 \be
 \label{EFP}
E = 3P = -3F
 \ee
holds.

One of the simple and instructive exercises which can be
now done is looking at the limit $N_c \to \infty$. We see
that the energy density becomes infinite in this limit. That
means that if we start to heat the system from $T =0$, we
just cannot reach the $QGP$ state --- to this end, an
infinite
energy should be supplied !

This physical conclusion can also be reached if looking at
the problem from the low temperature end. The spectrum of
$QCD$ in the limit $N_c \to \infty$ involves infinitely many
narrow states. The density of states grows exponentially
with energy
\footnote{One of the way to see it is to use the string
model for the hadron spectrum. A string state with large
mass is highly degenerate. The number of states with a given
mass depends on the number of ways $p(N)$ the large integer
$N \sim M^2/\sigma$ ($\sigma$ is the string tension) can be
decomposed in the sums of the form $N = \sum_i n_i$ (see
e.g. \cite{Polbook}). $p(N)$ grows exponentially with $N$.
That does not mean, of course, that in real $QCD$ with large
$N_c$ the spectrum would be also degenerate. Numerical
calculations in $QCD_2$ with adjoint matter fields show that
there is no trace of degeneracy and the spectrum displays a
stochastic behavior \cite{Kleb}. And that means in
particular that there is little hope to describe {\it
quantitatively} the $QCD$ spectrum in in the limit $N_c \to
\infty$ in the string model framework. But a qualitative
feature that the density of states grows exponentially as
the energy increases is common for the large $N_c$ $QCD$ and
for the string model.}
 \be
 \label{dens}
\rho(E) \propto e^{cE}
\ee
 That means that the partition function
 \be
 \label{Zdens}
 Z \sim \int \rho(E) e^{-E/T} dE
 \ee
is just not defined at $T \geq c^{-1}$. There is a Hagedorn
temperature $T_H = c^{-1}$ above which a system cannot be
heated \cite{Hagedorn}.

When $N_c$ is large but finite,  no limiting temperature
exists (It is seen also from the low temperature viewpoint:
at finite $N_c$ the states have finite width and starting
from some energy begin to overlap and cannot be treated as
independent degrees of freedom), and one can bring the
system to the $QGP$ state when supplying enough energy.
But when $N_c$ is large, the required energy is also large.
That suggests (though does not prove, of course) that at
large finite $N_c$ a first order phase transition with a
considerable latent heat takes place. This conjecture is
supported by the lattice measurements, but we will not
pursue this discussion further and refer an interested
reader to our review \cite{Varenna}.

\subsection{Debye screening.}

Consider the gluon polarization operator in $QGP$ with
account
of thermal loop corrections. It is transverse, $k_\mu
\Pi_{\mu\nu}(k) = 0$. At zero temperature, transversality
and Lorentz-invariance dictates the form $\Pi_{\mu\nu}(k) =
\Pi(k^2) (g_{\mu\nu} - k_\mu k_\nu /k^2)$. At finite $T$,
Lorentz-invariance is lost and the polarization operator
presents a combination of two different (transverse and
longitudinal) tensor structures. Generally. one can write
\be
\label{tens}
\Pi_{00} = \Pi_l(\omega, |\vec{k}|) \nonumber \\
\Pi_{i0} = \frac{k_i\omega}{|\vec{k}|^2} \Pi_l(\omega,
|\vec{k}|) \nonumber \\
\Pi_{ij} = - \Pi_t(\omega, |\vec{k}|) (\delta_{ij} - k_ik_j
/|\vec{k}|^2) + \frac{\omega^2}{\vec{k}^2} \frac {k_i
k_j}{\vec{k}^2} \Pi_l(\omega, |\vec{k}|)
\ee
Consider first the longitudinal part of the polarization
operator in the kinematic region where $\omega$ is set to
zero in the first place after which $k \equiv |\vec{k}|$ is
also sent
to zero. By the reasons which will be shortly seen, we
denote this quantity $m_D^2$:
  \be
  \label{mDdef}
m_D^2 = \lim_{k \to 0} \Pi_l(0, k)
 \ee
To understand the physical meaning of this quantity,
consider the correlator
  \be
\label{screen}
 <A_0^a(\vec{x}) A_0^b(0)> \sim \delta^{ab} \int \frac
{d\vec{k}}{\vec{k}^2 + m_D^2} e^{i\vec{k} \vec{x}} \propto
e^{-m_D |\vec{x}|}
 \ee
Thus, $m_D$ coincides with the inverse screening length of
chromoelectric potential $A_0$.

There is a clear analog with the usual plasma. A static
electric charge immersed in the plasma is screened by the
cloud of ions and electrons so that the potential falls down
exponentially $\propto \exp\{-r/r_D\}$ where the {\it Debye
radius} $r_D$ is given (in an ordinary non-relativistic
plasma) by the expression (see e.g. \cite{LP})
 \be
\label{rD}
r_D^{-2} = \frac {4\pi n e^2}T + \frac {4\pi n (Ze)^2}T
 \ee
where $n$ is the electron and ion density, $Z$ is the ion
charge (so that the second term describes the ion
contribution), and,  to
avoid unnecessary complications,
  we assumed that the electron and ion
components of the plasma have the same temperature $T$
Calculating the one-loop thermal contribution to the gluon
polarization operator (see Fig.1), one can easily obtain an
analogous
formula for $QGP$:
 \be
 \label{mD}
m_D^2 = r_D^{-2} = \frac {g^2 T^2}3 \left(N_c  + \frac
{N_f}2 \right)
 \ee
where the first term describes the screening due to thermal
gluons and the second term --- the screening due to thermal
quarks.
The result (\ref{mD}) was  first obtained
 by Shuryak \cite{ShurD}. It has the same
structure as (\ref{rD}) (Note that the density of particles
in ultrarelativistic plasma is expressed via the
temperature, $n \propto T^3$).
It is worth mentioning that the Debye screening is
essentially a classical effect and not only quarks but also
gluons result in screening rather than antiscreening. That
should be confronted with the famous antiscreening of the
charge in Yang-Mills theory at zero temperature due to
quantum effects.

\begin{figure}
\begin{center}
\SetScale{1.333}
\begin{picture}(200,50)(0,0)
\Photon(0,20)(10,20){1}{3}
\Photon(30,20)(40,20){1}{3}
\PhotonArc(20,20)(10,0,360){1}{10}
\put(13,0){a)}
\put(0,17){k}
\put(13,24){p}

\Photon(80,20)(120,20){1}{10}
\PhotonArc(100,30)(10,0,360){1}{10}
\put(67,0){b)}

\Photon(160,20)(170,20){1}{3}
\Photon(190,20)(200,20){1}{3}
\CArc(180,20)(10,0,360)

\Photon(240,20)(250,20){1}{3}
\Photon(270,20)(280,20){1}{3}
\DashCArc(260,20)(10,0,360){5}

\put(120,0){c)}
\put(173,0){d)}
\end{picture}

\vspace{0.5cm}
\caption{Gluon polarization operator in one loop.}
\end{center}
\end{figure}
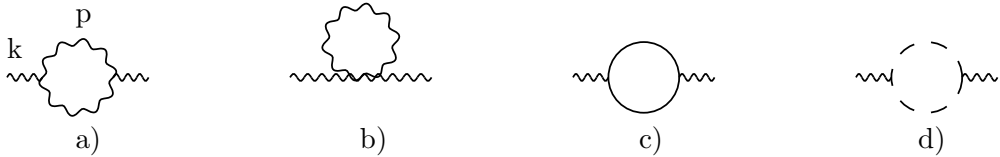

In $QED$, the correlator $<A_0(\vec{x}) A_0(0)>_T$ is a
gauge
invariant object and also the notion of charge screening is
unambiguous and well defined --- there {\it are} classical
electric charges and one can measure the potential of such a
charge immersed in plasma by standard classical devices. Not
so in QCD. We do not have at our disposal classical color
charges due to confinement and the experiment of measuring
the chromoelectric field of a test color charge cannot be
carried out even in principle. Also the gluon polarization
operator which enters the definition (\ref{screen}) of  a
Debye screening mass is generally speaking gauge-dependent.
Thus, a question of whether the Debye screening mass is a
physical notion in a non-abelian theory is fully legitimate.

 The answer to this question is positive. But with
reservations.

Note first of all that though the gluon polarization
operator is generally gauge-dependent, the Debye screening
mass defined in (\ref{mDdef}) does not depend on the gauge
{\it in the leading order} (we shall discuss also non-leading 
corrections in the next section). That suggests that
the result (\ref{mD}) has an invariant physical meaning.

Indeed, one can consider a gauge-invariant correlator
  \be
  \label{CP}
C(\vec{x}) = <P(\vec{x}) P^*(0)>_T
  \ee
where
  \be
  \label{Pdef}
P(\vec{x}) = \frac 1{N_c} {\rm Tr} \exp \left\{
i \int_0^\beta A_0(\vec{x}, \tau) d\tau \right\}
  \ee
is a gauge-invariant operator (the so called Polyakov line,
a special thermal version of the Wilson line \cite{Polsus}).
If $r = |\vec{x}|$ is not too large (the exact meaning of
this will be specified later), the  correlator has an
exponential behavior
  \be
  \label{CmDr}
C(r) \propto [G_{00}(r)]^2 \propto \exp\{-2m_Dr\}
  \ee
The correlator (\ref{CP}) can be attributed a physical
meaning. $-T \ln C(r)$ coincides with the change of free
energy of the system when putting there a pair of heavy
quark and antiquark at distance $r$, the averaging over
color spin orientations being assumed \cite{Larry}.

Also we shall see in the next section that the perturbative
corrections $\sim g^3T^4$ to a perfectly well defined and
physical quantity --- the free energy --- depend directly on
the value of $m_D$.

Thus, the notion of Debye screening mass in $QGP$ is
physical, though, unfortunately, not to the same extent as
it is in the usual plasma. The correlator (\ref{CP}) with
the correlation length of fractions of  a fermi cannot be
directly measured even in a {\it gedanken} laboratory
experiment --- at least, we cannot contemplate such an
experiment. But it can be measured in lattice numerical
experiments which is almost as good. We shall see later that
a number of other characteristics of $QGP$ have a similar
semi-physical status.

\subsection{Magnetic  screening.}
Debye mass describes the screening of static electric
fields. In abelian plasma, magnetic fields are not screened
whatsoever. Such a screening could be provided by magnetic
monopoles, but they are not abundant in Nature. The absence
of monopoles is technically related to one of the Maxwell
equations $\partial_i B_i = 0$. In the non-abelian case, the
corresponding equation reads ${\cal D}_i B_i = 0$ where
$\vec{\cal D}$ is a covariant derivative. Thus, gluon field
configurations with local color magnetic charge density
$\rho_m \sim \partial_i B_i$ (with usual derivative) are
admissible. The presence of such configurations in the gluon
heat bath results in screening of chromomagnetic fields.

Let us see how it comes out in a perturbative calculation.
Consider the one-loop graph in Fig.1a which contributes to
the polarization operator of the spatial components of the
gluon field $\Pi_{ij}(0, \vec{k}) \propto \Pi_t(0,
k)$. At high temperatures, it suffices to take into
account only the lowest Matsubara frequency (with $\omega
= 0$) in gluon propagators. We have
  \be
  \label{Pitr}
\Pi_t(0, k) \sim g^2 k^2 T \int \frac {d^3p}
{\vec{p}^2 (\vec{p} - \vec{k})^2}  \sim g^2 T k
  \ee
The numerical coefficient can be explicitly calculated, but
it depends on the gauge and makes as such a little sense.
The loop integral is determined by the low momentum region
$|\vec{p}|_{char} \sim k \ll T$ (as we are
interested in the large distance behavior of the gluon
Green's function, we keep $k$ small). Note that the
Feynman integral for $\Pi_l(0, k)$ has a completely
different behavior being saturated (in the leading order) by
the loop momenta $|\vec{p}| \sim T$ --- it is a so called
{\it hard thermal loop}.

The transverse part of the gluon Green's function is
\be
\label{Gtran}
 G_t(0, k) \sim \frac 1{\vec{k}^2 + \Pi_t(0, k)}
\ee
  We see that at $k \sim g^2 T$ the one-loop contribution to
the polarization operator is of the same order as the tree
term $k^2$. One can estimate a two-loop contribution to
$\Pi_t(0, k)$ which is of order $(g^2 T)^2$. The factor
$T^2$ here comes from two loops (we use the rule
(\ref{intT}) and take into account only the contribution of
the lowest Matsubara frequency in each loop).
Similarly, a three-loop contribution to $\Pi_t(0,
k)$ is of order $(g^2 T)^3/k$, a four-loop contribution is
of order $(g^2 T)^4/k^2$ etc. (The growing powers of $k$ in
the denominator are provided by infrared 3-dimensional loop
integrals. Infrared integrals  may also provide for a
logarithmic singularity in external momentum $k$ in the two-
loop
contribution which is of no concern for us here).
At $k \sim g^2T$ all contributions are of the same order and
the perturbation theory breaks down \cite{Linde}.

There is an alternative way to see it. Let us write down the
expression for the partition function of the pure glue
system at finite temperature:
  \be
  \label{ZglT}
Z = \int \prod dA^a_\mu (\vec{x}, \tau) \exp \left\{
- \frac 1{4g^2} \int_0^\beta d\tau \int d^3x
(F^a_{\mu\nu})^2 \right\}
  \ee
When $T$ is large and $\beta$ small, the Euclidean time
dependence of the fields may be disregarded. Also the
effects due to $A_0(\vec{x}, \tau)$ can be disregarded ---
time components of the gluon field acquire the large mass
$m_D \sim gT \gg g^2T$ and decouple. We are left with the
expression
  \be
  \label{Z3}
Z = \int \prod dA^a_i (\vec{x}) \exp \left\{
- \frac 1{4g^2T} \int d^3x (F^a_{ij})^2 \right\}
  \ee
A theory with quarks is also reduced to (\ref{Z3}) in this
limit --- the fermions have high Matsubara frequencies $\sim
T$ and decouple. The partition function (\ref{Z3}) describes
a non-linear 3D theory with the dimensional coupling
constant $g_3^2 \sim g^2T$
\footnote{One should be careful here. It would be wrong to
use the expression (\ref{Z3}) for calculating, say, the free
energy of $QGP$ at large $T$. The latter takes the
contributions from hard thermal loops (involving also
fermions!) with momenta of order $T$ which are not taken
into account in Eq.(\ref{Z3}). (\ref{Z3}) should be
understood as an {\it effective} theory describing soft
modes with momenta of order $g^2T$.}.
No perturbative calculations in this theory are possible.

In $QED$, there is no magnetic screening (and the effective
3-dimensional theory is trivial). Presumably, perturbative
calculations in hot $QED$ can be carried out at any order,
though this question is not {\it absolutely} clear.

What is the physical meaning of this non-abelian screening ?
Can it actually be measured ?

The way we derived it, the magnetic screening shows up in
the large distance behavior of the spatial gluon propagator
$G_t(r) \propto \exp\{ik^* r\}$ where $k^*$ is the solution
to the  equation
  \be
  \label{disp}
k^2 + \Pi_t(0, k) = 0
  \ee
  There is no reason to expect that $k=0$ is a solution ---
the behavior of $\Pi_t(0, k)$ in the limit $k \to 0$ where
the
pertubation theory does not work is not known, but one can
tentatively guess that it tends to a constant $\sim
(g^2T)^2$ (As we have seen, $g^2T$ is the only relevant
scale in this limit). If so, $k^* \sim ig^2T$ and $G_t(r)$
falls down
exponentially $\propto \exp\{-Cg^2Tr\}$ at large distances.
Hence the term ``magnetic screening''.

Unfortunately, the gluon polarization operator is a gauge-dependent
quantity and, would the God provide us with the
exact expression for $\Pi_t(0, k)$ in any gauge, even He
could not guarantee that the solution of the dispersive
equation (\ref{disp}) would be gauge-independent
\footnote{We shall return to the discussion of this
important point in the following sections.}.

What one can do, however, is to consider the correlator of
chromomagnetic fields $C_{ab}(\vec{x}) \ =\
<\vec{B}^a(\vec{x}) \vec{B}^b(0)>_T$.
This correlator also depends on the gauge in a non-abelian
theory, but the gauge dependence amounts to rotation in
color
space: $C_{ab}(\vec{x}) \to \Omega_{a'a}(\vec{x})
C^{a'b'}(\vec{x}) \Omega_{b'b}(0)$ and cannot affect the
exponential behavior of the propagator. This is the way the
magnetic mass is usually measured on the lattices.

One can do even better considering gauge-invariant
correlators, the simplest one is \ \ $<G_{\mu\nu}^2(\vec{x})
G_{\mu\nu}^2(0)>_T$. Then the quantity
  \be
  \label{mudef}
\mu = - \lim_{r \to \infty} \frac 1r \ln
<G_{\mu\nu}^2(\vec{x})
G_{\mu\nu}^2(0)>_T
  \ee
provides an invariant definition of the magnetic screening
mass. The indefinite article is crucial here. Choosing other
gauge-invariant correlators, one would get other invariant
definitions. For example (and this will be important in the
following discussion) , the true correlation length of the
correlator of Polyakov loops (\ref{CP}) is also of order
$(g^2T)^{-1}$. The asymptotics (\ref{CmDr}) is an {\it
intermediate} one and holds only in the region
  \be
\label{range}
  (gT)^{-1} \ll r \ll (g^2T)^{-1}
  \ee

What was wrong with our previous derivation ? The matter is
we took into account earlier only the thermal corrections to
the gluon propagator and tacitly neglected corrections to
the vertices. Ab accurate analysis \cite{Nad} shows that
this is justified in the range (\ref{range}) but not beyond.
An example of the graph providing the leading contribution
in $<P(\vec{x}) P^*(0)>_T$ at $r \gg (g^2T)^{-1}$ is given
in Fig.2.
\footnote{
The  asymtotics of the correlator of Polyakov loops at large
enough $r$ is determined by the magnetic photon exchange
also in abelian plasma. Two magnetic photons can be coupled
to $P(x)$ via a fermion loop. In abelian case, magnetic
photons are massless and, as a result, the correlator has a
power asymptotics $\propto 1/r^6$ \cite{Arnold}. Physically,
it corresponds to Van-der-Vaals repulsion between the clouds
of virtual  electrons and positrons formed near two heavy probe 
charges.
Seemingly, the asymptotics $\propto 1/r^6$ for the Polyakov
lines correlator in $QGP$ found in \cite{Wong} has this
origin. But in non-abelian case, the asymptotics becomes
exponential when taking into account the magnetic screening
effects.}

\begin{figure}
\begin{center}
\SetScale{2}
\begin{picture}(200,60)(0,0)
\Vertex(50,30){1}
\Vertex(150,30){1}

\PhotonArc(65,30)(15,0,360){1}{10}
\PhotonArc(135,30)(15,0,360){1}{10}

\DashCArc(100,-4.4)(40,60,120){3}
\DashCArc(100,64.4)(40,240,300){3}
\end{picture}
\end{center}
\caption{}
\end{figure}

Thus, the ``experimental status'' of the magnetic mass
(\ref{mudef}) and of other similar quantities is roughly the
same and even better than for the Debye mass. The
exponential behavior $\propto \exp\{-\mu r\}$ of the
correlators is expected to hold at any large $r$ and the
value $\mu$ can in principle be determined in lattice
experiments with any desired accuracy (now the accuracy is
very poor due to a finite size of available lattices, but it
is not a question under discussion here). On the other hand,
the Debye screening mass cannot be determined with an
arbitrary accuracy due to the finite range (\ref{range})
where the
asymptotics (\ref{CmDr}) holds (being sophisticated enough,
an invariant
definition of Debye mass still can be suggested for $N_c \geq
3$ --- see the
discussion in the following section). And, as was also the
case
for the Debye mass, we cannot invent any laboratory
experiment where the magnetic mass could be directly
measured.

\section{Static Properties of $QGP$: Perturbative
Corrections.}
\setcounter{equation}0

\subsection{Debye mass.}

To provide a smooth continuation of the discussion started
at the end of the previous section, we consider first 
higher-loop effects in the Debye mass. Note first of all that the
definition (\ref{mDdef}) is not suitable anymore. 
Higher-loop corrections $\Delta m_D \sim g^2T$ to the Debye mass as
defined in Eq.(\ref{mDdef}) depend on the gauge
\cite{Kaj}. A better way is to define the Debye mass as
the solution of the dispersive equation $k^2 + \Pi_l(0,k) =
0$. In other words, we define \cite{Rebhan}
 \be
 \label{mDReb}
m_D^2 = \lim_{k^2 \to -m_D^2} \Pi_l(0,k)
  \ee
The longitudinal part of the gluon Green's function
$$
G_l(0,k) = \frac 1{k^2 + \Pi_l(0,k)}
$$
has then the pole at $k = im_D$. It is conceivable that the
pole
position is gauge-invariant even though the Green's function
itself is not. (See, however, a discussion in the following
section. There {\it are} cases when formal arguments
displaying gauge-invariance of the pole position fail in
higher orders of perturbation theory due to severe infrared
singilarities. It is not clear for us whether they really
work in the case of Debye mass.)
  In the next-to-leading order, the (gauge-invariant)
result is \cite{Rebhan}
  \be
  \label {mDcorr}
\frac {\Delta m_D^2}{m_D^2} = \frac {N_c}{2\pi}
\sqrt{\frac 3{N_c + N_f/2}} g \ln \frac 1g
  \ee
We see that the correction is non-analytic in coupling
constant. The non-analyticity appears due to bad infrared
behavior of the loop integrals --- they involve a
logarithmic infrared divergence and depend on the low
momenta cutoff which is of order of magnetic mass scale
$g^2T$. Thus, we have
$$
\Delta m_D^2 \propto \int_{\mu_{mag}}^{m_D} \frac {dp}p
= \ln {m_D \over \mu_{mag}} \sim \ln {1 \over g}
$$
The magnetic infrared cutoff is actually provided by higher-order
graphs --- the orders of perturbation theory are mixed
up and a pure two-loop calculation is not self-consistent.
As we have seen, $\mu_{mag}$ cannon be determined
analytically which means that the correction $\sim g^2T$
without the logarithmic factor in the Debye mass cannot be
determined analytically.

Also the correction $\sim g^2T$  cannot be ``experimentally
observed''. Really, we have seen that the invariant physical
definition of $m_D$ refers to the correlator of Polyakov
loops (\ref{CP}). The latter displays the exponential
behavior (\ref{CmDr}) in the limited range (\ref{range}).
But that is tantamount to saying that the correction $\sim
g^2T$ in the Debye mass cannot be determined from the
correlator (\ref{CP}). To do this, one should probe the
distances $r \gg (g^2T)^{-1}$ where the correlator has a
completely different behavior being determined by the
magnetic scale. The correction (\ref{mDcorr}) is still
observable, however, due to a logarithmic enhancement
factor.

To be more precise, the correlator in the range
(\ref{range}) has the form \cite{Nad}
 \be
  \label{CPcorr}
<P(\vec{x}) P^*(0)>_T \propto \exp \left\{
-2m_D^{(0)} r \left [1 + \frac {N_c}{2\pi}
\sqrt {\frac 3{N_c + N_f/2}} g \ln(m_D^{(0)}r) \right]
\right\}
 \ee
where $m_D^{(0)}$ is the lowest order Debye mass
(\ref{mD}). At finite $r \ll (g^2T)^{-1}$, the infrared
cutoff in the loop integrals is provided by $r^{-1}$ rather
than $\mu_{mag}$ and we have
$$ \ln {m_D \over \mu_{mag}} \to \ln (m_Dr) $$
The Fourier image of the correlator (\ref{CPcorr}) does {\it not}
have a singularity at finite $k$ whatsoever. Thus, the Debye mass pole
in the gluon propagator does not show up as a pole in the
gauge-invariant correlator of Polyakov loops.

But, anyway, in the theoretical limit $g(T) \to 0$ there is
a range of $r$  where the condition $r \ll (g^2T)^{-1}$ is
fulfilled so that the intermediate asymptotic law
(\ref{CPcorr}) holds whereas the correction in the exponent
$\propto gm_D^{(0)} r \ln (m_D r)$ is large compared to 1
and can be singled out in numerical lattice experiment.

There are two interesting recent proposals to define Debye
mass non-perturbatively in a gauge-invariant way
\cite{Arnold}. The first one is to consider the correlator
of imaginary parts of Polyakov lines in a complex
representation which is invariant under the action of the
center of the group $Z_N$ (say, the decouplet representation
in $SU(3)$)
\footnote{The authors of Ref.\cite{Arnold} argued the
necessity to consider only the
$Z_N$ - invariant representations saying that otherwise the
correlators
give zero after averaging over different ``$Z_N$ - phases''.
Actually, such
phases do not exist in Nature and a proper agrumentation
should be that the
correlators which are not invariant under $Z_N$
transformations just do not have
a physical meaning \cite{bub}.}.
The point is that $P^*_{10} - P_{10}$ is odd
under Euclidean time reversion and cannot be coupled to
magnetic gluons. That means that the correlator
 \be
 \label{CP10}
<P^*_{10}(\vec{x}) - P_{10}(\vec{x}),\ \ P^*_{10}(0) -
P_{10}(0)>_T
 \ee
exhibits the Debye screening falloff $\propto \exp\{-3m_D
r\}$ even at arbitrary large $r$. This definition does not
work, however, for $SU(2)$ where the Polyakov line is real
in any representation.

Another suggestion was to study the behavior of large spatial
Wilson loops in {\it adjoint} color representation. At large
distances, the theory is effectively reduced to the 3-
dimensional
YM theory (\ref{Z3}). Adjoint color charges in
this theory are not confined but rather screened, and the
Wilson loop exhibits the perimeter law behavior
  \be
  \label{W3per}
W(C) \propto \exp \{ -m^* \times {\rm perimeter}(C)\}
\ee
One can identify $m^*$ (which is of order $g^2T$) with a
non-perturbative correction to the Debye mass. The problem
here that $m^*$ has no trace of the lowest order
contribution (\ref{mD}). It involves a part of the
perturbative correction ( \ref{mDcorr}) $\sim g^2T \ln \frac
1g$
due to gluon loops but not a similar contribution from
fermion loops. Still $m^*$ certainly has an invariant
meaning and is as such an interesting quantity to study.

\subsection{Free energy.}
This is the most basic and physical quantity of all. May be
this is the reason why perturbative corrections are known
here with record precision.

\begin{figure}
\begin{center}
\SetScale{2}
\begin{picture}(200,60)(0,0)

\PhotonArc(15,30)(15,0,360){1}{10}
\PhotonArc(45,30)(15,0,360){1}{10}

\DashCArc(100,30)(25,0,360){4}
\Photon(100,5)(100,55){1}{7}

\CArc(170,30)(25,0,360)
\Photon(170,5)(170,55){1}{7}
\end{picture}
\end{center}
\caption{ Free energy in two loops.}
\end{figure}
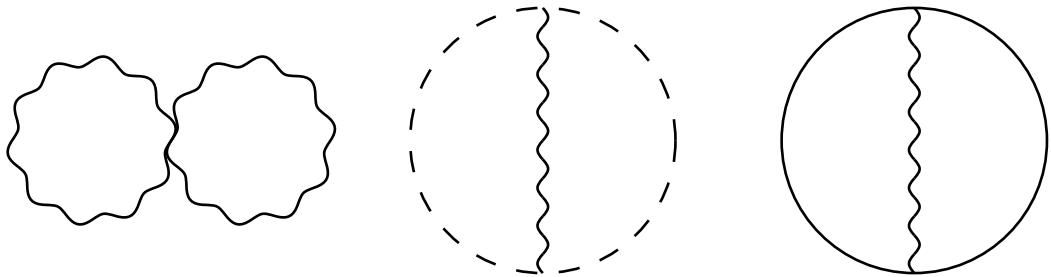

The correction $\sim g^2 T^4$ has been found by Shuryak
\cite{ShurD}. To this end, one should calculate the two-loop
graphs depicted in Fig.3. The behavior of Feynman integrals
in this order is quite benign and no particular problem
arises.

\begin{figure}
\begin{center}
\SetScale{2}
\begin{picture}(200,80)(0,0)

\PhotonArc(100,30)(30,0,360){1}{10}
\Photon(85,4)(85,56){1}{7}
\Photon(115,4)(115,56){1}{7}
\put(100,63){k}

\end{picture}
\end{center}

\caption{ An infrared-divergent 3-loop graph in free
energy.}
\end{figure}
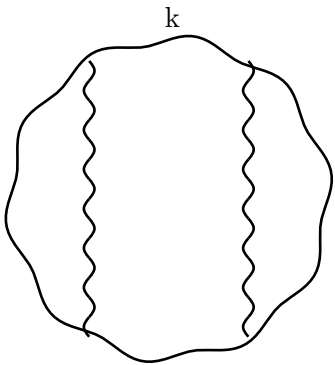

However, on the three-loop level, the problem  crops up. The
contribution of the graph depicted in Fig.4 is infrared
divergent:
  \be
  \label{F5}
\Delta F^{Fig.4} \sim g^4 T^5 \int \frac{dk}{k^2}
 \ee
That means that a pure 3-loop calculation is not self-
consistent and, to get a finite answer, one has to resum a
set of infrared-divergent graphs in all orders of
perturbation theory. This is, however, not a hopeless
problem, and it has been solved by Kapusta back in 1979
\cite{Kapusta}. The leading infrared singularity is due to
the so called ``ring diagrams'' depicted in Fig.5.

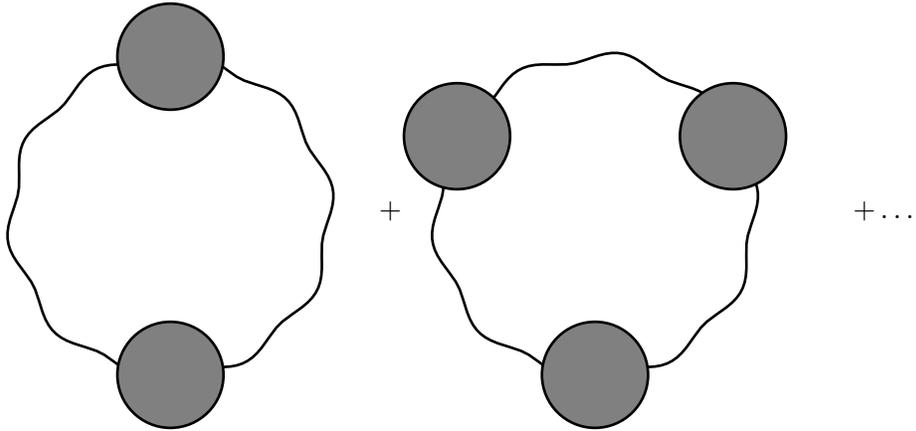
\begin{figure}
\begin{center}
\SetScale{2}
\begin{picture}(200,100)(0,0)

\PhotonArc(50,50)(30,0,360){1}{10}
\GCirc(50,20){10}{0.5}
\GCirc(50,80){10}{0.5}

\PhotonArc(130,50)(30,0,360){1}{10}
\GCirc(130,20){10}{0.5}
\GCirc(104,65){10}{0.5}
\GCirc(156,65){10}{0.5}

\put(90,50){+}
\put(180,50){$+\ldots$}

\end{picture}

\end{center}
\caption{ Ring diagrams. Grey circles stand for a set of
one-loop graphs in Fig.1.}
\end{figure}

The sum of
the whole set of ring diagrams has the form
  \be
  \label{Fring}
F^{ring} \sim T \int d^3k \left\{ \ln\left[1 +\frac
{\Pi_l(0,k)}{k^2} \right] -  \frac {\Pi_l(0,k)}{k^2}
\right\}
\ee
The expansion of the integrand in $
{\Pi_l(0,k)}/{k^2}$ restores the original infrared-divergent
integrals. However, the whole integral is convergent being
saturated by momenta of order $k \sim [\Pi_l(0,0)]^{1/2}
\sim gT$. Neglecting the $k$-dependence in the polarization
operator, we obtain
   \be
  \label{F3}
F_{3} \sim T \int d^3k \left[ \ln\left(1 +\frac
{m_D^2}{k^2} \right) -  \frac {m_D^2}{k^2} \right] \sim
Tm_D^3 \sim g^3T^4
\ee
We see that the correction is non-analytic in the coupling
constant $\alpha_s$  which exactly reflects the fact that the individual
graphs diverge and the orders of perturbation theory are
mixed up. Note, however, that infrared divergences here are
of comparatively benign variety --- the integral depends on
the scale $k^{char} \sim gT$ and we are far from the land of
no return $k \sim g^2T$. That is why an analytic
determination of the coefficient in (\ref{F3}) is possible.

The next correction has the order $\sim g^4 T^4 \ln(1/g)$.
It comes from the same ring graphs of Fig.5 where now the
term $\sim g^2Tk$ in $\Pi_l(0,k)$ should be taken into
account. This correction was determined by Toimela
\cite{Toimela}.

Presently, the correction $\sim g^4T^4$ (without the
logarithmic factor)
\cite{Zhai1} and the
correction $\sim g^5T^4$ \cite{Zhai2,Nieto} are known. This
is the absolute limit beyond which no perturbative
calculation is possible --- similar ring graphs as in Fig.5
but with magnetic gluons would give the contribution
$\propto T\mu_{mag}^3 \sim g^6 T^4$ in the free energy. And,
as we have seen, $\mu_{mag}$ cannot be determined
perturbatively.

Collecting all the terms, the following result is obtained
\cite{Zhai2,Nieto}
  \be
 \label{Ffull}
F = - \frac{8\pi^2T^4}{45}
\left[ F_0 + F_2 \frac {\alpha_s(\mu)}{\pi} + F_3 \left(
\frac {\alpha_s(\mu)}{\pi} \right)^{3/2} +
\right. \nonumber \\
\left.
F_4 \left( \frac {\alpha_s}{\pi} \right)^{2} +
F_5 \left( \frac {\alpha_s}{\pi} \right)^{5/2} +
O(\alpha_s^3 \ln \alpha_s ) \right]
 \ee
where
\be
\label{F0c}
F_0 = 1 + \frac {21} {32} N_f
 \ee

\be
\label{F2c}
F_2 = -\frac{15}4 \left(1 + \frac {5} {12} N_f \right)
 \ee

\be
\label{F3c}
F_3 = 30 \left( 1 + \frac {N_f}6 \right)^{3/2}
 \ee

\be
\label{F4c}
F_4 = 237.2 + 15.97 N_f - 0.413 N_f^2 + \frac {135}2
\left( 1 + \frac {N_f}6 \right) \ln \left[ \frac
{\alpha_s}\pi \left( 1 + \frac {N_f}6 \right) \right]
\nonumber \\
- \frac {165}8 \left(1 + \frac {5} {12} N_f \right)
\left(1 -\frac {2} {33} N_f \right) \ln \frac \mu{2\pi T}
 \ee

 \be
 \label{F5c}
F_5 = \left(1 + \frac { N_f}6 \right)^{1/2} \left[
-799.2 - 21.96 N_f - 1.926 N_f^2 \right. \nonumber \\
\left. + \frac {485}2 \left(1 + \frac { N_f}6 \right)
\left(1 -\frac {2} {33} N_f \right) \ln \frac \mu{2\pi T}
\right]
 \ee
The expressions (\ref{F0c}) - (\ref{F5c}) are written for
$N_c =3$. The coefficients like 237.2 are not a result of
numerical integration but are expressed via certain special
functions.

A nice feature of the result (\ref{Ffull}) is its renorm-invariance.
The coefficients $F_4$ and $F_5$ involve a
logarithmic $\mu$--dependence in such a way that the whole
sum does not depend on the renormalization scale $\mu$.

Let us choose $\mu = 2\pi T$ (this is a natural choice,
$2\pi T$ being the lowest non-zero gluon Matsubara frequency ).
In that case, we have
 \be
 \label{series}
F = F_0 \left[1 - 0.9 \alpha_s + 3.3\alpha_s^{3/2} + (7.1 + 3.5
\ln \alpha_s ) \alpha_s^2 - 20.8 \alpha_s^{5/2} \right]
 \ee
Note a large numerical coefficient at $\alpha_s^{5/2}$. It
is rather troublesome because the correction $\sim
\alpha_s^{5/2}$ overshoots all previous terms up to very
high temperatures and, at temperatures which can  be
realistically ever reached at accelerators, makes    the
whole perturbative approach problematic.

Take $T \sim 0.5 \ {\rm GeV}$ (this is the temperature one can
hope to
achieve at RHIC \cite{RHICT}). Then $2\pi T \sim 3 \ {\rm
GeV}$
and $\alpha_s \sim 0.2$. (We use a conservative estimate for
$\alpha_s$ following from $\Upsilon$ physics \cite{Vol}.
Recent
measurements at LEP favor even larger values.). The series
(\ref{series}) takes the form
  \be
  \label{sernum}
 F = F_0[1 - .18 + .3 +.06 - .37 + \ldots]
 \ee
which is rather unsatisfactory. We emphasize that the
coefficient of $\alpha_s^{5/2}$ is rather trustworthy being
obtained independently by two different groups.

 The last remark is technical. The result (\ref{Ffull}) was
obtained in Euclidean technique. There is a real time
calculation which correctly reproduces the two-loop term
$\sim g^2T^4$ in free energy \cite{Landsman}, but nobody so
far succeeded in calculating in this way the terms $\sim g^3
T^4$ and higher. Certainly, real-time technique is not very
suitable for calculation of static quantities, and one way
to get the result is good enough, but, to my mind, it is an
interesting methodical problem.

\section{Collective excitations.}
\setcounter{equation}0
One of striking and distinct physical phenomena
characteristic of usual plasma is a non-trivial dispersive
behavior of electromagnetic waves. In contrast to the vacuum
case where only transverse photons with the dispersive law
$\omega = |\vec{k}|$ propagate, two different branches
with different non-trivial dispersive laws $\omega_\bot(k)$
and $\omega_\|(k)$ appear in plasma. The value
$\omega_\bot(0) = \omega_\|(0) = \omega_{pl}$ characterizes
the
eigenfrequency of spatially homogeneous charge density
oscillations and is called the plasma frequency.

A similar phenomenon exists also in $QGP$. The spectrum of
$QGP$ involves collective excitations with quantum number of
quarks and gluons. Like in usual plasma, there are
transverse and longitudinal branches of gluon collective
excitations (alias, transverse and longitudinal {\it
plasmons}) and their properties {\it on the one-loop level}
are very similar to the properties of photon collective
excitations in usual plasma. A novel feature is the
appearance of non-trivial fermion collective excitations
({\it plasminos}).
But, again, they are not specific for a non-abelian theory
and appear also in ultrarelativistic $e^+e^-$ plasma (in the
limit when the electron mass can be neglected compared to
the temperature-induced gap in the electron spectrum
$\propto eT$).

 \subsection{One-loop calculations.}
The dispersive laws of quark and gluon collective
excitations can be obtained via solution of a non-abelian
analog of the Vlasov system involving the classical field
equations in the medium and Boltzmann kinetic equation
\cite{Blaizot}.
\footnote{To find quark dispersive laws, one has to write
down generalized kinetic equations for spinor densities.
Such equations were first introduced in \cite{sound} when
studying the problem of goldstino dispersion law in a
supersymmetric thermal medium.}
This way of derivation makes analogies with usual plasma
(where the Vlasov system is a standard technique) the most
transparent.

I will outline here another way of derivation which is more
conventional and more easy to understand for a field
theorist. This is actually the way the results were
originally derived \cite{Kal,Weldon}.

Consider the gluon Green's function in a thermal medium. As
was mentioned in sect. 2, at $T \neq 0$, different kinds of
Green's function exist. To be precise, we are considering
now the {\it retarded} Green's function
  \be
  \label{Gret}
\left[ G_{\mu\nu}^{ab} (x) \right]^R = i<\theta(t)
[A_\mu^a(x), A_\nu^b(0)] >_T
  \ee
which describes a response of the system on a small
perturbation applied at $t = 0$ at some later time $t > 0$.
\footnote{As in commonly used gauges $G^{ab} \propto
\delta^{ab} $, we shall suppress color indices in the
following. We have retained them here just to make clear
that we are dealing with a commutator of Heisenberg field
operators, not with a commutator of classical color fields.}
The Fourier image of (\ref{Gret}) is free of singularities
in the upper $\omega$ half--plane. The poles of
$G^R_{\mu\nu}(\omega, \vec{k})$ correspond to eigenmodes of
the system and exactly give us the desired spectrum of gluon
collective excitations. The dispersive equation
 \be
\label{disptens}
\det \| G^R_{\mu\nu} (\omega, k) \| = 0
 \ee
splits up in two:
  \be
  \label{displt}
\omega^2 - k^2 - \Pi_t(\omega, k) = 0 \nonumber \\
k^2 + \Pi_l(\omega, k) = 0
\ee
with $\Pi_{l,t}(\omega, k)$ being defined in
Eq.(\ref{tens}). The solutions to the equations
(\ref{displt}) give two branches of the spectrum.

The explicit one-loop expressions for $\Pi_l(\omega, k)$ and
$\Pi_t(\omega, k)$ obtained by the calculation of the graphs
in Fig.1 in the limit $\omega, k \ll T$ are
\cite{Kal,Weldon}
  \be
  \label{Pilt}
\Pi_l(\omega, k) = 3\omega_{pl}^2 [1 -  F(\omega/k)]
\nonumber \\
\Pi_t(\omega, k) = \frac 32 \omega_{pl}^2 \left[ \frac
{\omega^2}{k^2} + \frac{k^2 - \omega^2}{k^2} F(\omega/k)
\right]
 \ee
where
 \be
\label{ompl}
\omega_{pl} = \frac{gT}3 \sqrt{N_c + \frac {N_f}2}
 \ee
is the plasma frequency and
 \be
\label{Fx}
F(x) = \frac x2 \ln \frac{x+1}{x-1}
 \ee

\begin{figure}
\grpicture{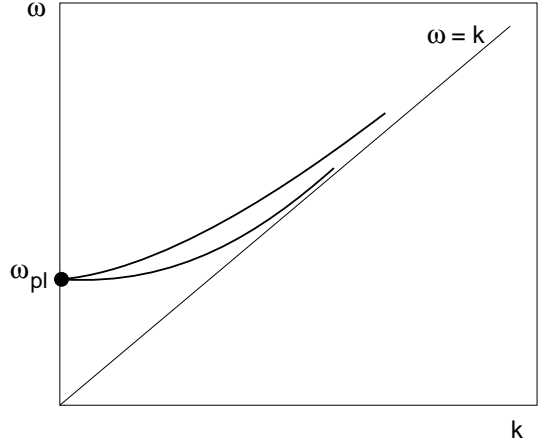}
\caption{ Plasmon spectrum in one loop.}
\end{figure}

 The behavior of dispersive curves is schematically shown in Fig.6. At $k
= 0$, $\omega_\bot(0) = \omega_\|(0) = \omega_{pl}$. Then
the two branches diverge:
  \be
\label{lowk}
 \omega^2_\bot(k \ll gT) = \omega_{pl}^2 + \frac 65 k^2
\nonumber \\
 \omega^2_\|(k \ll gT) = \omega_{pl}^2 + \frac 35 k^2
\ee
At $k \gg gT$ both branches tend to the vacuum dispersive
law
  \be
\label{highk}
 \omega^2_\bot(k \gg gT) = k^2 + \frac 32 \omega_{pl}^2
\nonumber \\
 \omega^2_\|(k \gg gT) = k^2 \left[1 + 4\exp \left(
 - \frac{2k^2}{3\omega_{pl}^2} - 2 \right)
 \right]
\ee
We see that $\omega_\|(k)$ approaches the line $\omega = k$
exponentially fast.

The dispersive laws for quark collective excitations are
obtained from a similar analysis of the quark Green's
function. At $T \neq 0$, the fermion polarization operator
involves two tensor structures $\omega \gamma^0 - \vec{k}
\vec{\gamma}$ and $\omega \gamma^0 + \vec{k} \vec{\gamma}$
which gives rise to two dispersive branches. We will call
the branch corresponding to the Lorentz-invariant structure
``transverse'' and the branch corresponding to the structure
$\omega \gamma^0 + \vec{k} \vec{\gamma}$ ---
``longitudinal''. These terms may be misleading in the
fermion case because, in contrast to the plasmons with
photon or gluon quantum numbers, these branches are not
associated with transverse and longitudinal field
polarizations. Hence the quotation marks. But better names
were not invented, and using the words ``transverse'' and
``longitudinal''  still makes a certain sense because the
physical properties of ``transverse'' and ``longitudinal''
fermion branches are rather analogous to the physical
properties  of transverse and longitudinal gluon branches.

\begin{figure}
\grpicture{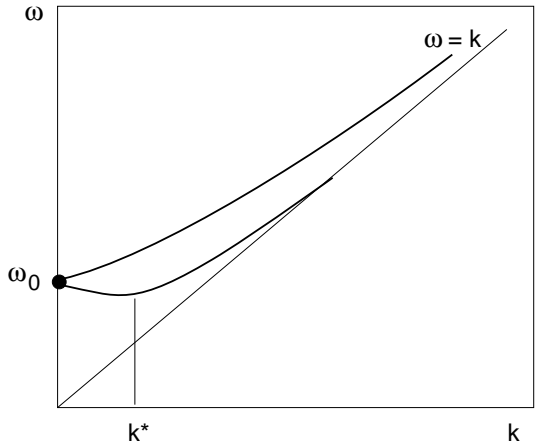}
\caption{ Plasmino spectrum in one loop.}
\end{figure}

%\begin{center}
%{\bf Fig..}
%\end{center}

%\vspace{3mm}
The pattern of the quark spectrum is shown in Fig.7. It is
similar to the gluon spectrum with one important distinction
--- at $k \sim 0$, $\omega_\bot(k)$ and  $\omega_\|(k)$
involve linear terms of opposite sign:
 \be
 \label{qlowk}
 \omega^q_\bot(k \ll gT) = \omega_0 + \frac k{3}
\nonumber \\
 \omega^q_\|(k \ll gT) = \omega_0 - \frac k{3}
 \ee
where $\omega_0$ is the plasmino frequency at zero momentum
:
\be
\label{om0}
\omega_0^2 = \frac{g^2T^2}8 c_F
 \ee
Thus,  $\omega^q_\|(k)$ first goes down and reaches minimum
at
some $k^*$. The group velocity of the longitudinal plasmino
at
this point is zero.

\subsection{Landau Damping.}
The quoted one-loop results for the dispersive laws of
transverse plasmons
and plasminos are gauge-invariant and stable with respect to
higher-order
corrections. The latter is not true, however, for
longitudinal excitation
branches \cite{Silin,spectr}. We have seen that
$\omega_\|^{1\ loop}(k)$ tends
to  the line $\omega=k$ exponentially fast at $k \gg gT$.
This can be easily
seen from the analysis of the dispersive equations for
longitudinal branches
which in the limit $k \gg gT$ have the form
  \be
  \label{displk}
\omega_\| + k \sim \frac{g^2T^2}k \ln \frac k{\omega_\| -k}
 \ee
At $k \gg gT$, the solution exists when the logarithm is
large and
$\omega_\| - k$ is exponentially small. The logarithmic
factor
in Eq.(\ref{displk}) comes from the angular integral
 \be
 \label{ang}
\sim \int \frac{d\theta}{\omega - k \cos \theta}
 \ee
which diverges at $\omega = k$. This collinear divergence
appears due to
masslessness of quarks and gluons in the loop depicted in
Fig.1. But
 quarks and gluons in $QGP$ are {\it not} massless --- their
dispersive law
acquires the gap $\sim gT$ due to temperature effects. An
accurate calculation
requires substituting in the loops the {\it dressed}
propagators. As a result,
the logarithmic divergence in the integral (\ref{ang}) is
cut off and the
logarithmic factor in Eq.(\ref{displk}) is modified:
  \be
  \label{cutoff}
\ln \frac k{\omega - k} \to \frac 12 \ln \frac{k^2}{(\omega
- k)^2 + Cg^2T^2}
\ee
Dressing of propagators amounts to going beyond one-loop
approximation.
Strictly speaking, to be self-consistent one should also
take into account
one-loop thermal corrections to the vertices (this procedure
is known as
{\it resummation of hard thermal loops} \cite{Pisar}), but
in this particular case these
corrections do not play an important role. What {\it is}
important is the
cutoff of the logarithmic collinear singularity due to
effective
temperature-induced masses.

Substituting (\ref{cutoff}) in (\ref{displk}), we see that
the new
dispersive equation does not at all have solutions with real
$\omega$ for
large enough $k$. This fact can be given a natural physical
explanation.
When $k$ is small compared to $gT$, there is no logarithmic factor
in the dispersive equation, the modification
(\ref{cutoff}) is
irrelevant, and the dispersive law of longitudinal modes does
not deviate
from the one-loop result. Then the logarithm appears, the modification
(\ref{cutoff}) starts
playing a role, and, at some $k^{**} \sim gT$, the longitudinal
dispersive
curve crosses the line $\omega = k$. At this point the
longitudinal
polarization operator acquires the imaginary part due to
{\it Landau
damping}.

In usual plasma, Landau damping is the process when
propagating
electromagnetic waves are ``absorbed'' by the electrons
moving
in plasma. In
the language of quantum field theory, it is a $2 \to 1$
process
  \be
\label{21}
\gamma^* + e \to e
  \ee
In real time technique, that corresponds to a contribution
to the imaginary
part of the polarization operator so that both internal
electron lines in
the loop are placed on mass shell. At $T = 0$ the standard
Cutkovsky rules
imply positive energies of all particles in the direct
channel, and
the imaginary part appears only due to the decay $\gamma^* \to
e^+ + e^-$. At
$T \neq 0$, Cutkovsky rules are modified and both signs for
energy are
admissible. Physically, that corresponds to the presence of
real particles
in the heat bath so that the process (\ref{21}) may go.

Also in $QGP$ imaginary parts of polarization operators may
acquire
contributions due to Landau damping. The corresponding
processes are
 \be
  \label{12qg}
g^* + g \to g, \ \ g^* + q \to q,\ \ q^* + g \to q,\ \ q^* +
\bar{q}
\to g, \ \ldots
 \ee
where $g^*, q^*$ are plasmon and plasmino collective
excitations and $q, g$
are the excitations with characteristic momenta of order of
temperature
(in this kinematic region, the dispersive laws are roughly
the same as for
tree quarks and gluons and the star superscript is
redundant).

The kinematic condition for the scattering processes
(\ref{12qg}) to go is
that the frequency of collective excitations $g^*$ and $q^*$
would be less
than their momentum. We have seen that the condition $\omega
< k$ is
realized indeed for longitudinal plasmon and plasmino
excitations starting
from some  $k^{**} \sim gT$. At $k > k^{**}$, the Landau
damping switches on
and the dispersion law acquires an imaginary part. The
imaginary part
rapidly grows and very soon becomes of order of the real
part. From there
on it makes no sense to talk about propagating longitudinal
modes anymore.
The situation is the same as in usual plasma where
longitudinal modes also
become overdamped at large enough momenta and disappear from
the physical
spectrum \cite{LP}.

\begin{figure}
\grpicture{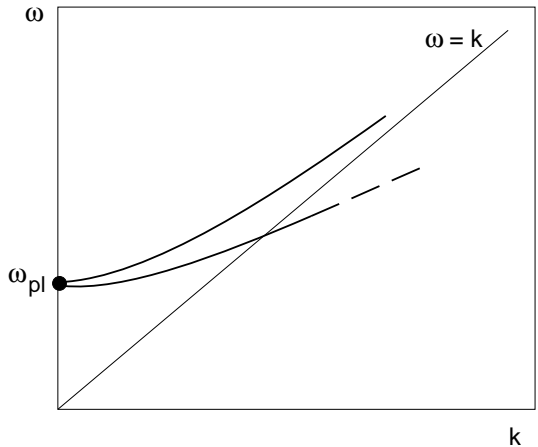}
\caption{ Plasmon spectrum.}
\end{figure}

The true pattern of plasmon collective modes is shown in
Fig.8. A similar
picture holds for plasminos.

\subsection{Observability.}

One-loop dispersive curves are gauge-invariant. However, the
question
whether these curves are physically observable is, again,
highly non-trivial. It is easy to measure explicitly photon dispersive
curves in usual
plasma --- to this end, one should study the propagating
classical
electromagnetic waves, measure the electric charge density
(say, by laser
beams) as a function of time and spatial coordinates and
determine thereby
the frequency and the wave vector of the wave.

But there is no such thing in Nature as classical gluon
field due to
confinement, and no classical device which would measure the
color charge
density exists. Even more obviously, quark fields (which
have Grassmannian
nature) cannot be treated classically. Hence, one cannot
really measure the
energy and momentum of propagating colored waves in a
direct physical
experiment.

What one {\it can} measure are correlators of colorless (in
the first place
, electromagnetic) currents. Modification of dispersive laws
affects these
correlators and that can be observed. However, a colorless
current always
couples to a {\it pair} of  colored particles and, as a
result, physical
correlators involve some integrals of quark and gluon
Green's functions
which are related to quark and gluon dispersive
characteristics only in an
indirect way. Also, thermal modification of vertices is
 as important here
as modification of the Green's functions.

However, there is one special point on the dispersive curves
which can in
principle be directly measured in experiment. This is the
point on the
longitudinal plasmino curve where its frequency acquires a
minimal value
and the group velocity turns to zero.
Consider the problem of
emission of relatively soft $e^+ e^-$ or $\mu^+\mu^-$ pairs
by $QGP$.
In a ``thermos bottle''
experiment, one should make sure that the size of your
thermos bottle is
much less than the lepton mean free path. Otherwise, the
leptons are
thermalized and their spectrum is just Planckian. But in
heavy ion collisions  experiments, $QGP$ is produced in
small volume, the condition $L_{\rm char} \ll L^{em}_{\rm free\ path}$ 
is satisfied,
and the spectrum of emitted leptons (and photons) can provide a
non-trivial
information on dynamic characteristics of $QGP$.

The spectrum of soft dileptons was calculated in \cite{Yuan}.
This is one
of very few  {\it physical} problems we know of
where the hard
thermal loop resummation
technique \cite{Pisar} should be used (and was used) at full
length. The spectrum feels the
effects due to quark and gluon interactions in the region
$E_{l^+l^-} \sim
P_{l^+l^-} \sim gT$ ---  the spectrum at larger energies and momenta
is the same as for the gas
of free quarks. One particular source of soft dileptons is
the process
$q^*_\bot \to q^*_\| \ +\ l^+l^-$. The probability of this
process has a
``spike'' for the momentum  of $q_\bot^*$ and $q_\|^*$
coinciding with the
momentum $k^*$ on the longitudinal plasmino dispersive curve
with zero group
velocity. There are just many plasminos at the
vicinity of
this point and the phase space factor provides a singularity
at $E_{l^+l^-}
= \omega_\bot(k^*) - \omega_\|(k^*)$ in the spectrum.
Another spike comes
from the process when a longitudinal plasmino with momentum
$k^*$
annihilates with a longitudinal antiplasmino with the opposite
 momentum to produce a lepton pair with the energy $2\omega_\|(k^*)$.

Unfortunately, in the soft region, the main contribution in
the spectrum is
due to cuts. In other words, the most relevant elementary kinetic
processes are
not  $q^* \to q^* + l^+l^-$ or $q^* + \bar{q}^* \to
l^+l^-$, but rather
$q^* + g^* \to q^* + g^* + l^+l^-$ etc. The spikes actually
have finite
width due to collisional damping of collective excitations
(the issue to be
discussed in the next section), and one can hope to see only
a tiny
resonance on a huge background. Still, such a resonance in
the spectrum
{\it is } an observable effect.

\section{ Damping Mayhem and Transport Paradise.}
\setcounter{equation}0

\subsection{Direct decay.}
In the previous section, we discussed the Landau damping
contribution to
the imaginary parts of polarization operators and,
correspondingly, to
imaginary parts of dispersion laws. It comes from the
kinematic region
$\omega < k$ and is physically related to absorbtion of
ingoing excitations
by thermal quanta like in (\ref{12qg}). However, we did not
say a word about the
contribution of direct decay processes $g^* \to q + \bar{q}$
etc. in the
timelike kinematic region $\omega > k$.

That was with a good reason. On the one-loop level, the
contribution of
decay processes in imaginary parts is nonzero and is of
order $\sim g^2T$.
Unfortunately, it depends on the gauge and, in some gauges,
has even the
wrong sign corresponding not to damping of excitations but
to instabilities
\cite{Zahed}. The point is that such one-loop calculation is
unstable with
respect to higher-order corrections. It is very clear
physically --- quarks
and gluons in $QGP$ cannot be treated as massless but
acquire dynamical
masses due to thermal effects. And the decay of a plasmon or
plasmino into
two other collective excitations is not kinematically
allowed. The only
exception is the process $g^* \to q^* + \bar{q}^*$ which in
principle may
go if \cite{spectr}
 \be
 \label{Nf6}
N_f > 9c_F - c_V = 6
 \ee
But there are at most three light flavors in real $QGP$ and
decay processes
can be safely forgotten.

\subsection{Collisional damping.}
Still, damping is there even in the timelike region $\omega
> k$ due to
collisions $g^* + q^* \to g^* + q^*$, $q^* + \bar{q}^* \to
g^* + g^*$ etc.
This is also the main source of damping of transverse
electromagnetic waves
in usual plasma \cite{LP}. A rough estimate for collisional
damping in $QGP$
can be  done very simply.

The meaning of damping is the inverse lifetime of excitations.
We have
\be
\label{dampest}
\zeta \sim (\tau_{life})^{-1} \sim n \sigma^{tot}
  \ee
where $n \sim T^3$ is the density of the medium and
$\sigma^{tot}$ for
excitations which carry (color) charge has a Coulomb form
  \be
  \label{sigma}
\sigma^{tot} \sim g^4 \int \frac{dp_\bot^2}{p_\bot^4} \sim
\frac {g^2}{T^2}
 \ee
We took into account the fact that the power infrared
divergence for the
integral of Coulomb cross section is effectively cut off at
$p_\bot \sim
gT$ due to Debye screening
\footnote{and due to Landau damping effects --- see more
detailed discussion below.}. As a result, we obtain the
estimate
\be
\label{g2T}
\zeta^{q,g} \sim g^2T
 \ee

Note that this value for the damping is unusually  large. It
is much larger
than, say, the damping of photons in ultrarelativistic
$e^+e^-$ -- plasma.
The latter can also be estimated from the formula
(\ref{dampest}), but
$\sigma^{tot}$ is now not the Coulomb, but the Compton cross
section. The
integral has now the form $\int_{eT} dp_\bot^2/p_\bot^2 \sim
\ln (1/e)$ and
the estimate is
  \be
  \label{zetgam}
\zeta^\gamma \sim   e^4T \ln (1/e) \ll e^2T
 \ee
The question arises whether the new anomalously large scale
$\sim g^2T$ has a \ physical \ relevance \footnote{Do not
confuse this scale with the magnetic scale
which is also
of order $g^2T$. The former is related to kinetic properties
of the system
while the latter refers exclusively to static phenomena.}.
 We will return to discussion of this point a bit later.

\begin{figure}
\begin{center}
\SetScale{2}
\begin{picture}(200,70)(0,0)
\Line(55,35)(70,35)
\Line(130,35)(145,35)

\PhotonArc(100,35)(30,0,180){1}{10}
\CArc(100,35)(30,180,360)

\Vertex(100,5){3}
\Vertex(100,65){3}

\put(55,38){k}

\end{picture}

\end{center}
\caption{}
\end{figure}

 An accurate calculation of the damping of {\it fast moving} ($k \gg
gT$) quark and gluon excitations in
$QGP$ has been done in \cite{spectr,andam} (see also
\cite{Alt}).
Consider the graph in Fig.9
for the quark polarization operator where the lines with blobs
stand for quark and
gluon propagator dressed by thermal loops.
\footnote{It can be shown \cite{spectr} that when
calculating the leading
contribution in $\zeta$ in
the kinematic region $k \gg gT$ which is under discussion
now, vertex
corrections can be disregarded.}
 The imaginary part of the whole loop in Fig.9 depends on
the imaginary
parts of internal propagators. The imaginary part of the
gluon propagator
due to Landau damping turns out to be of paramount
importance. Physically,
this contribution just corresponds to the scattering
processes
$g^* + q^* \to g^* + q^*$ and $g^* + g^* \to g^* + g^*$
as can be
easily inferred if spelling out the exact gluon propagator
as in Fig.10 (there is also a similar graph with internal gluon loop).

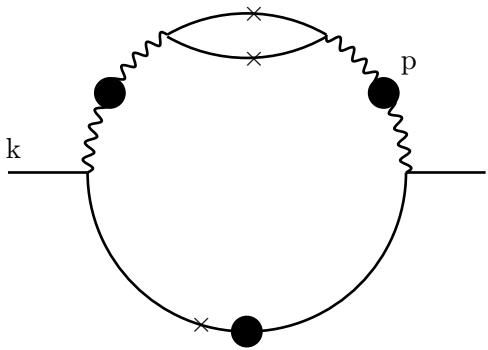
\begin{figure}
\begin{center}
\SetScale{2}
\begin{picture}(200,70)(0,0)
\Line(55,35)(70,35)
\Line(130,35)(145,35)

\PhotonArc(100,35)(30,0,60){1}{10}
\PhotonArc(100,35)(30,120,180){1}{10}
\CArc(100,35)(30,60,120)
\CArc(100,86.6)(30,240,300)

\CArc(100,35)(30,180,360)

\put(100,64){$\times$}
\put(100,55.5){$\times$}
\put(90,5){$\times$}

\Vertex(100,5){3}
\Vertex(125.8,50){3}
\Vertex(74.2,50){3}

\put(55,38){k}
\put(130,55){p}
\end{picture}

\end{center}
\caption{ A graph contributing to $ {\rm Im}\ 
\Sigma(k)$.  Crossed lines are put on mass shell. One of
the quarks in the internal loop has a negative energy.}
\end{figure}

 Dressing  the quark propagator (indicated by the blob in
Fig.10) is  important. If not
taking it
into account, the imaginary part of the propagator is 
$\delta$ --
function and the result for ${\rm Im}\ \Sigma(k)$ has the
form
  \be
  \label{dkdk0}
 {\rm Im}\ \Sigma(k) = - \frac 3{32\pi} \omega_{pl}^2 g^2 c_F
T \gamma^0
\int dp_0 \delta(p_0 - p \cos \theta) \int_{-1}^1 d \cos
\theta
\int \frac {dp^2}{p^4 + \frac 9{16} \omega_{pl}^4 \frac
{p_0^2}{p^2}}
  \ee
The integral has a power infrared behavior at $p \gg gT$, but this divergence
is cut off due to Landau damping effects [the second term in the denominator
in the RHS of Eq.(\ref{dkdk0})].
Still, the integral in (\ref{dkdk0}) diverges logarithmically at $p
\ll gT$ and the result for the damping inferred from Eq.(\ref{dkdk0})
is infinite.

The crucial observation is that the {\it dressed} quark propagator does not 
have singularities on the real $p_0$
axis. A self-consistent account of the collisional damping
for the quark
Green's function moves its singularities in the complex
plane.
As a result, $\delta$ -- function in the integrand is
replaced by a smooth
distribution with the width of order $\zeta \sim g^2T$. This
smoothing cuts
off the logarithmic singularity in (\ref{dkdk0}) at $p \sim
g^2T$. The
other source for the cutoff could be provided by magnetic
screening effects
, but the latter is an essentially non-abelian phenomenon
whereas the
cutoff due to smearing out the $\delta$ -- function is a
universal effect
which occurs also in an abelian theory.

To find the dispersion law, one should add Eq.(\ref{dkdk0})
to the one-loop
result for $\Sigma(k)$ and solve the dispersive equation.
The solution is
complex $\omega^{pole}(k) = \omega^{1\ loop}(k) - i\zeta
(k)$
  \footnote{A refined analysis which takes into account the
modification
of ${\rm Im} \ \Sigma (k)$ for complex $\omega$ when one starts to move
from the real $\omega$ axis towards a singularity and which is
beyond the scope of
this lecture
shows that the dispersive equation has actually no solutions
and the
singularity is no longer a pole, but a branching point
\cite{BNN,damp1,Pilon}. But this branching point is located
at the same distance from the real axis as the would-be pole
and brings about
the same damping behavior  of the gluon retarded Green's
function
$G^R(t) \sim \exp\{-\zeta t\}$ at large real times. 
There is a recent claim \cite{nonexp} that in abelian theory
$G^R(\omega)$ does not involve
singularities at all at finite distance from the real axis. This would
result in a non-exponential
decay of  $G^R(t)$ at large real time --- $G^R(t) \sim \exp\{- \alpha
 T t\ \ln(eTt)\}$  \  (cf. Eq.(\ref{CPcorr}). We do not 
think, however, that it is correct --- such a form of $G^R(t)$ does not 
conform with a smooth behavior of $G^R(\omega)$ on the {\it real} $\omega$ axis
where our calculation of $\Sigma(k)$ is well under control and the cutoff
due to a finite fermion width {\it should} be taken into account.}
  and the final result for $\zeta(k \gg gT)$ is very simple.
 \be
 \label{dampLS}
\zeta^q = \alpha_s c_F T \ln(C^q/g)  \nonumber \\
\zeta^g = \alpha_s c_V T \ln(C^g/g)
\ee
 Only the coefficient of logarithm can be calculated. The
constants $C^q,\ C^g$ under
the logarithm  cannot determined. Actually, we will see shortly that these
constants are gauge--dependent and cannot be {\it defined} in a reasonable
way.

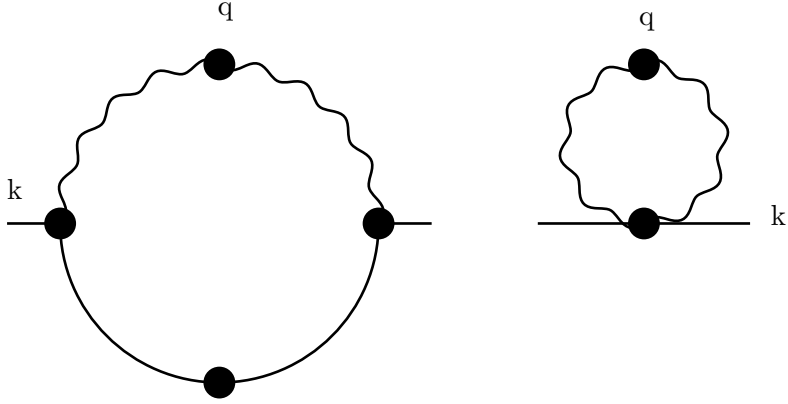
\begin{figure}
\begin{center}
\SetScale{2}
\begin{picture}(200,70)(0,0)

\put(0,40){k}
\Line(0,35)(10,35)
\Line(70,35)(80,35)

\put(40,75){q}
\PhotonArc(40,35)(30,0,180){1}{10}
\CArc(40,35)(30,180,360)

\Vertex(40,5){3}
\Vertex(40,65){3}
\Vertex(10,35){3}
\Vertex(70,35){3}

\put(145,35){k}
\Line(100,35)(140,35)
\PhotonArc(120,50)(15,0,360){1}{10}
\put(120,73){q}
\Vertex(120,65){3}
\Vertex(120,35){3}

\end{picture}

\end{center}
\caption{Polarization operator of soft plasmino.}
\end{figure}

Damping of excitations in another kinematic region $\vec{k}
= 0$ (``standing''
plasmons and plasminos) was studied in \cite{Pisar}. Here
the vertex
corrections are as important as corrections to the
propagators, however an
accurate analysis of \cite{Pisar} shows that it suffices to
take into
account only one (hard thermal) loop corrections both in
polarization operators
and  vertices. Consider for definiteness the damping of
standing plasminos.
The graphs contributing to the soft plasmino polarization
operator are
shown in Fig.11. Using Keldysh technique one can derive in
the limit of
soft external momentum
  \be
  \label{SigmaR}
\Sigma^R(k) = -2ig^2 \int \frac {d^4q}{2\pi^4} \frac T q_0
{\rm Im}\ 
D_{\mu\nu}^R(q) \left[ \frac 12 \Gamma^R_{\mu\nu}(k;q) +
\right.
\nonumber \\
\left. \Gamma^R_\mu (k,q+k; q) G^R(q+k) \Gamma_\nu^R(q+k,k;-
q) \right]
\ee
($G^R$ and $D^R$ are the retarded Green's functions and
$\Gamma^R$ are the retarded
vertices).
Subsituting here transverse and longitudinal parts of the
gluon Green's
function and one-loop vertices $\Gamma_{\mu\nu}$ and
$\Gamma_\mu$,
calculating the integral (which in this case can be done
only numerically),
and solving the dispersive equation, one arrives at the
result
 \be
 \label{dampPis}
\zeta^q(k=0) = 1.43 c_F \frac {g^2T}{4\pi}
 \ee
However, the gluon Green's function involves also a 
gauge-dependent part
  \be
  \label{Dgauge}
D_{\mu\nu}^{R(\alpha)}(q) = (\alpha -1) \frac{q_\mu q_\nu}{[(q_0 +
i\epsilon)^2 - q^2]}
 \ee
where $\alpha$ is a gauge parameter and an infinitesimal
$i\epsilon$ is
introduced to provide for the right analytical properties.
At  first
sight, this gauge-dependent piece should not affect the
position of the
pole. Really, one can use the Ward identities (which hold
also at finite
temperature for {\it retarded} propagators and vertices
\cite{Taylor,andam})
to derive
  \be
\label{SigWard}
\Sigma^{R(\alpha)} (k) = -ig^2 T c_F (\alpha -1) [G^R(k)]^{-
1}
\int \frac {d^4q}{2\pi^4} \frac 1  q_0 \left\{ \frac
{1}{[(q_0 +
i\epsilon)^2 - q^2]} - \frac{1}{[(q_0 - i\epsilon)^2 - q^2]}
\right\}
\nonumber \\
\left[ G^R(k+q) - G^R(k) \right] [G^R(k)]^{-1}
  \ee
We see  the presence of factors $[G^R(k)]^{-1}$ both on the
right and on the left. $G^R(k)$ is singular at the pole and
$[G^R(k)]^{-1}$ is zero. One might infer from this that a
gauge-dependent contribution to $\Sigma^R({\rm  pole})$ and
hence to the corresponding solution of the dispersive
equation determining the pole position is also zero.

However, this is wrong \cite{BKS,damp1}. The point is that
the integral in (\ref{SigWard}) involves a severe power
infrared divergence and is infinite at the pole. We have a 
thereby $0 \times \infty$ uncertainty. This uncertainty can
be resolved by
 choosing $k$ not exactly at the pole but slightly off mass
shell.
\footnote{To determine the exponential asymptotics of
$G^R(t)$ at large real times, we have to stay on the real
$\omega$ axis i.e. {\it off mass shell} \cite{damp1}.}
 Then $[G^R(k)]^{-1}$ is not exactly zero and also the
divergence in the integral is cut off  by an 
off-mass-shellness. When the distance from the mass shell is small, the
final result for $\Sigma^{R(\alpha)}(k \to {\rm pole})$ does
not depent on this distance and is just finite. In the soft
momenta region
 \be
 \label{Siggauge}
\Sigma^{R(\alpha)}_{\rm soft}(k) = - \frac
{i(\alpha -1) g^2 T c_F}{4\pi} \gamma^0
 \ee
This brings about a gauge-dependent part $\sim g^2T$ in the
damping of soft plasminos. A similar analysis with the same
conslusion can be carried out for plasmons. 

The same
gauge-dependence shows up in the damping of energetic
plasmons and
plasminos, but in the latter case, this gauge-dependence is
parametrically overwhelmed by
the leading gauge-independent contribution (\ref{dampLS})
involving the factor $\ln (1/g)$.

\subsection{Observability.}
The observed gauge-dependence of the damping obviously
indicates that it is not a physical quantity. This is
definitely true at least for soft plasmons and plasminos 
where the gauge-dependent part and the gauge-independent part
(\ref{dampPis}) are of the same order $\sim g^2T$.

Indeed, it is not possible to contemplate a physical
experiment where this quantity could be measured.
That should be confronted with the case of abelian plasma
where damping of electromagnetic waves is a perfectly
physical quantity and can be directly observed by measuring
the attenuation of the amplitude of a classical wave with
time. But as we already noted, no classical gluon or quark
waves exist. This observation refers also for damping of
{\it electron and positron} collective excitations in the
ultrarelativistic  abelian plasma. It also has an
anomalously large scale $\sim e^2T$ (with the extra
logarithmic factor \ $\ln (1/e)$ for $k \gg eT$) and it also
cannot be directly measured.

One could try to observe the effects due to damping in
gauge-invariant quantites like the polarization operator of
electromagnetic currents. An accurate analysis which goes
beyond the conventional hard thermal loop resummation
technique and effectively resums a set of ladder graphs
shows, however, that a self-consistent account of the
corrections due to damping in the quark Green's functions
and in
the vertices results in the exact cancellation of the
anomalously large scale $\sim g^2T$ in the final answer
\cite{andam}.
(for a similar analysis with a similar conclusion in scalar
QED see \cite{Rebscal}).

However, there is a physical problem where the  scale
$\sim g^2T$ can in principle show up.
This is the already discussed problem of lepton pair
production in $QGP$. We have seen that the spectrum of
leptons pairs invoves spikes associated with a special point
with zero group velocity on the longitudinal plasmino
dispersive curve.
Going beyond the hard thermal loop aproximation and taking
into account the effects due to collisional damping in the
Green's functions {\it and} in the vertices would bring
about a finite width for these spikes of order $g^2T$ and
there is a principle possibility to measure this width. This
problem has not been studied, and it is not clear by now
whether the width of the spike can by calculated
analytically and whether one can single out this spike out
of the background.

What one can say quite definitely is that this width
crucially depends on modification of vertices due to
collisional effects and has nothing to do with the
(gauge-dependent) position of the pole (or whatever  the
real singularity is \cite{BNN,damp1,Pilon}) of the quark and
gluon Green's function.

Thus, we are convinced that the latter is not a physical
quantity probably even for energetic plasmons in spite
of the fact that the
leading contribution (\ref{dampLS}) is gauge-independent there.
We just do not know how on Earth this quantity could be
measured.

\subsection{Transport Phenomena.}
 There is a lot of kinetic phenomena in $QGP$ which are
physical and measurable.  Indeed, nothing {\it in principle}
prevents measuring the electric resistance of a vessel with
$QGP$ or studying the flow of $QGP$ through narrow tubes.
They do not depend, however, on the anomalous damping scale
$g^2T$, but rather on a much smaller scale
  \be
 \label{char}
(\tau_{char})^{-1} \sim g^4T \ln(1/g)
  \ee
 This scale already appeared
in (\ref{zetgam}) determining the damping of electromagnetic
waves in $e^+e^-$ -- plasma. And it is also the scale which 
determines the
mentioned physical effects of viscousity and electric
conductivity, and  many others --- heat conductivity,
energy losses of a heavy particle moving through plasma, etc.

The appearance of the scale $g^4T \ln(1/g)$ has a clear
physical origin. All the mentioned effects are inherently
 related to the rate of relaxation of the system to thermal
equilibrium. The latter can be estimated as
  \be
  \label{relax}
(\tau_{rel})^{-1} \sim n \sigma^{trans}
 \ee
It looks the same as the estimate for lifetime
(\ref{dampest}) but with an essential difference --- in
contrast to (\ref{dampest}), the estimate (\ref{relax}) involves the
{\it transport} rather than the total cross section. The
transport cross--section is defined as
  \be
  \label{trans}
   \sigma^{trans} = \int d \sigma (1 - \cos \theta)
  \ee
where $\theta$ is the scattering angle. The factor $(1 -
\cos \theta)$ takes care of the fact that small--angle
scattering though contributes to the total cross section,
does not essentially affect the distribution functions
$n_g(\vec{p})$, $n_q(\vec{p})$ and is not effective in
relaxation processes. For the Coulomb scattering in
ultrarelativistic plasma, the transport cross section is
    \be
  \label{sigtr}
\sigma^{trans} \sim g^4 \int_{(gT)^2} \frac{dp_\bot^2}{p_\bot^4}
\frac {p_\bot^2}{T^2} \sim \frac {g^4}{T^2} \ln \frac 1g
 \ee
Multiplying it by $n \sim T^3$ and substituting it in
(\ref{relax}), the estimate (\ref{char}) is reproduced.

Viscousity and all other similar quantities can be
calculated analytically  in the leading order (probably,
magnetic infrared divergences prevent an analytic evaluation
of these quantities in next orders in $g$, but this question
is not yet well studied). It is interesting that Feynman
diagram technique proves to be technically unconvenient here
, and the good old Boltzmann kinetic equation is the tool
people usually use (see e.g. \cite{Baym}).

  Let us make two illustrative estimates which make clear
how the relaxation scale (\ref{char}) depending on the
transport cross--section (\ref{sigtr}) arises.

First, let us estimate the electric conductivity of $QGP$ 
(it is the quite conventional conductivity,  not the 
``color conductivity''
which is sometimes discussed in the literature, depends on the 
anomalous damping scale $g^2T$, and {\it is} not a physically 
observable quantity --- we do not have batteries with color
charge at our disposal). Suppose at $t=0$ the system was at
thermal equilibrium so that the quark distribution functions
are $n_0(\vec{p}, \vec{x}, t=0) = n_F(|\vec{p}|)$. 
When we switch on the electric field, the distribution function
starts to evolve according to the kinetic equation
  \be
  \label{kineq}
  \frac {\partial n(\vec{p}, \vec{x}, t)}{\partial t}  + \vec{v} 
  \frac {\partial n(\vec{p}, \vec{x}, t)}{\partial \vec{x}}
  = e\vec{E} \frac{\partial n 
  (\vec{p}, \vec{x}, t)}{\partial \vec{p}} + \ldots
  \ee
  Dots in RHS of Eq.(\ref{kineq}) stand for the collision term which becomes
  relevant at $t \sim x \sim \tau_{\rm rel}  \sim \tau_{\rm free\ path} 
  \sim [g^4 T \ln(1/g)]^{-1}$.
  Thus, the electric field brings about distortions of the distribution 
  function which grow up to the characteristic value
       $$\delta n \sim e  \vec{E} \frac {\vec{v}}T n_0 \tau_{\rm free \ path}
  \sim \frac {e  \vec{E} \vec{v}  T}{g^4 \ln(1/g)}   $$
  At this point, collisional effects stop the growth (a particle drifting
  in external electric field collides with a particle in the medium, forgets 
  what happened before, and starts drifting anew).
  The density of electric current in the medium is
   \be
    \label{j}
   \vec{j}  = e  \int \vec{v} \delta n \ d^3p
   \sim \vec{E} \frac {e^2 T}{g^4 \ln(1/g)}
   \ee
   The coefficient between $\vec{j}$ and $\vec{E}$ gives the conductivity. 
    
Let us estimate now the energy losses of a heavy energetic
quark in $QGP$. Of course, free quarks do not exist, but a
physical experimental setup would be sending into the bottle
with $QGP$ a heavy meson ${Q}\bar{q}$ with open beauty or
top. In $QGP$, the meson dissociates, and a naked
heavy quark propagates losing its energy due to interaction
with the medium. It goes out then   on the other side of the
bottle dressed again with light quarks, but not necessarily
in the same way as before.  When $M_Q \gg \Lambda_{QCD}$,
this dressing does not essentially affect its energy. A
heavy particle containing $Q$ can be detected and its energy
can be measured.

Suppose a heavy quark is ultrarelativistic, but its energy
is not high enough for the Cerenkov radiation processes to
be important. Then the energy would be lost mainly due to
individual incoherent scatterings.
  The mean energy loss in each scattering is $\Delta E \sim
T$ ($T$ --- is a characteristic energy of the particles in
heat bath on which our heavy quark scatters). The mean time
interval between scatterings is $\tau_{\rm free\ path} \sim
 [g^4T \ln (1/g)]^{-1}$. We obtain
  \be
  \label{dEdx}
-  \frac {dE}{dx} \sim \frac {\Delta E}{\tau_{\rm free\
path}} = C \alpha_s^2 T^2 \ln(1/g)
  \ee
This estimate turns out to be  correct up to the argument of the 
logarithm which in reality is energy--dependent \cite{Thoma}.
 The  numerical coefficient $C = 4\pi c_F/3$ was
determined by Bjorken \cite{Bjorken}

\section{Acknowledgements.}
I am indebted to J.P. Blaizot, S. Peigne, E. Pilon, A. Rebhan, and 
E. Shuryak for useful discussions, remarks and references. This work
has been done under the partial support of INTAS Grants
CRNS--CT93--0023, 93--283, and 94--2851.

\end{document}